\documentclass[]{emulateapj}

\newcommand{\bmgc}{B_{\mbox{\tiny \sc MGC}}}

\shorttitle{MGC: mergers by asymmetry and dynamics}
\shortauthors{De~Propris et al.}

\begin{document}


\title{The Millennium Galaxy Catalogue: The connection between close pairs and asymmetry;  
       implications for the galaxy merger rate}


\author{
Roberto~De~Propris,\altaffilmark{1}
Christopher~J.~Conselice,\altaffilmark{2}
Jochen~Liske,\altaffilmark{3}
Simon~P.~Driver,\altaffilmark{4}\\
David~R.~Patton,\altaffilmark{5}
Alister~W.~Graham,\altaffilmark{6} and
Paul~D.~Allen\altaffilmark{4}
}

\altaffiltext{1}{Cerro Tololo Inter-American Observatory, Casilla 603,
La Serena, Chile}
\altaffiltext{2}{School of Physics and Astronomy, University of
Nottingham, University Park, NG7 2RD, United Kingdom}
\altaffiltext{3}{European Southern Observatory,
Karl-Schwarzschild-Stra{\ss}e 2, 85748 Garching b.~M{\"u}nchen,
Germany}
\altaffiltext{4}{Scottish Universities Physics Alliance (SUPA), 
School of Physics and Astronomy, University of St.~Andrews, 
North Haugh, St.~Andrews, Fife, KY16 9SS, United Kingdom}
\altaffiltext{5}{Department of Physics and Astronomy, Trent
University, 1600 West Bank Drive, Peterborough, ON, K9J 7B8, Canada}
\altaffiltext{6}{Centre for Astrophysics and Supercomputing, Swinburne
University of Technology, Hawthorn, VIC 3122, Australia}


\begin{abstract}

We compare the use of galaxy asymmetry and pair proximity 
for measuring galaxy merger fractions and rates for a volume limited 
sample of $3184$ galaxies with $-21 < M_B -5 \log h < -18$ mag. and 
$0.010 < z < 0.123$ drawn from the Millennium Galaxy Catalogue.
Our findings are that:\\
(i) Galaxies in close pairs are generally more asymmetric than isolated 
galaxies and the degree of asymmetry increases for closer pairs. At least
35$\%$ of close pairs (with projected separation of less than 20 $h^{-1}$
kpc and velocity difference of less than 500 km $s^{-1}$) show significant
asymmetry and are therefore likely to be physically bound.\\
(ii) Among asymmetric galaxies, we find that at least $80\%$ are either
interacting systems or merger remnants. However, a significant fraction
of galaxies initially identified as asymmetric are contaminated by nearby
stars or are fragmented by the source extraction algorithm. Merger rates
calculated via asymmetry indices need careful attention in order to remove
the above sources of contamination, but are very reliable once this is
carried out.\\
(iii) Close pairs and asymmetries represent two complementary methods of 
measuring the merger rate. Galaxies in close pairs identify future mergers,
occurring within the dynamical friction timescale, while asymmetries are
sensitive to the immediate pre-merger phase and identify remnants.\\
(iv) The merger fraction derived via the close pair fraction and asymmetries
is about $2\%$ for a merger rate of $(5.2 \pm 1.0) \times 10^{-4}\ h^3$ 
Mpc$^{-3}$ Gyr$^{-1}$. 
These results are marginally consistent with theoretical simulations (depending 
on the merger time-scale), but imply a flat evolution of the merger rate with 
redshift up to $z \sim 1$.

\end{abstract}


\section{Introduction}

In the $\Lambda$-dominated Cold Dark Matter (CDM) model, galaxies form 
via a process of hierarchical merging, in which more
massive objects are assembled gradually via mergers of increasingly
more massive subunits at progressively lower redshifts (see
\citealt{baugh06} for an introduction).  Observations of the merger
rate of galaxies and its evolution with redshift provide important
benchmarks for comparison with theoretical predictions, according to
which massive galaxies have undergone several mergers between $z \sim
3$ and the present epoch, doubling their mass over the last half of
the Hubble time \citep{murali02,maller05}.

Unfortunately, it is generally difficult to ascertain observationally
if two galaxies are going to merge in the near future, or if and when
a galaxy last experienced a merger. Galaxies undergoing mergers or
interactions may exhibit peculiar morphological disturbances, such as
tidal tails \citep{arp66, toomre72,toomre77}, while merger remnants
may also show relic structures from their previous interactions, such
as shells or ripples \citep{malin80,malin83}. A more quantitative
version of this approach measures the asymmetry of a galaxy's light
distribution to identify interacting systems and merger remnants and
estimate the galaxy merger rate, e.g.\ using the CAS system (Bershady, 
Jangren \& Conselice 2000, \citealt{conselice03,conselice03b} and 
references therein).

Before galaxies merge they will be distinct systems with some radial
velocity difference and projected separation on the sky from which a
merger time-scale can be calculated based on dynamical friction
arguments. Galaxy pair statistics therefore could, in principle, provide
a powerful probe of mergers and their frequency. With modern imaging
and spectroscopic surveys it is now possible to select pairs of
galaxies which are not only close on the sky, but also at very similar
redshifts, and are therefore more likely to be gravitationally
bound. Such objects are referred to as `dynamically close pairs'. The
mathematical formalism to measure the close pair fraction and to
derive the merger rate from it was developed by \cite{patton00} for
the Second Southern Sky Redshift Survey and \cite{patton02} for the
Canadian Network for Observational Cosmology 2 (CNOC2) survey and
later applied to DEEP2 \citep{lin04} and the MGC \citep{depropris05}. 
Both of these latter studies found a low local merger rate, suggesting 
that only a small fraction of massive galaxies are formed by major 
mergers at $z < 1$.

There are therefore two methods for observationally measuring the
merger history -- finding galaxies in pairs that will eventually
merge, and locating highly asymmetric galaxies. Neither of these two
approaches is free from ambiguity however. For example, not all close
galaxy pairs will be dynamically bound and result in mergers. The 
observational problem is that it is impossible to constrain the projected 
relative motion of the two members of a close pair or to precisely determine 
the radial distance between them. \cite{patton00} estimate that close pair
samples may suffer from contamination by as many as $50$\% unphysical
pairs (superpositions). 

On the other hand, not all asymmetric galaxies are imminent or recent major 
mergers, as it is possible to induce significant asymmetries by other means 
(e.g., interactions, star formation, minor mergers, etc. -- see, e.g.,
Conselice, Bershady \& Jangren 2000a). Care must therefore be taken when 
selecting mergers based on asymmetry \citep{conselice03}.

The two techniques -- asymmetry and close pairs -- are also likely to probe
different stages of a merger: \cite{hernandez05} suggest that galaxy
morphology is very robust, and that significant asymmetries are
induced only by a close interaction. Furthermore, N-body simulations
indicate that a merger remnant may remain significantly asymmetric for
an extended period after the merger event \citep{conselice06}. Hence
asymmetry is likely to probe the immediate pre-merger and an extended
post-merger phase. In contrast, galaxy pairs will select all stages of
the pre-merger phase except the very last, where the two galaxies can
no longer be identified unambiguously as individual objects in the
imaging data.

It is important to understand how galaxy pairs and asymmetric galaxies
measure the merger process because, for observational reasons, the two
methods are preferentially employed in different redshift regimes. At
high redshift, highly complete spectroscopic samples are difficult to
obtain whereas relatively high resolution imaging is available from
HST. Hence selection by asymmetry is often the method of choice in
this regime \citep[e.g.][]{conselice03b}. At low redshift, the
situation is essentially reversed so that the pair method is preferred
\citep[e.g.][]{patton02}. We clearly require a cross-calibration of the 
two methods on the same dataset before drawing
reliable conclusions regarding the evolution of the merger rate at high
and low redshifts.

In this paper we attempt to deliver this cross-calibration by
searching for dynamically close pairs of galaxies and measuring the
structural asymmetries on the same sample of objects. We then
determine if paired galaxies are asymmetric, and also how the merger
fractions and rates derived from both methods compare. A few questions
we address include: are galaxies in close pairs more asymmetric than
isolated galaxies? Does the degree of disturbance correlate with pair
separation? Are more asymmetric galaxies more likely to lie in pairs,
with the most asymmetric ones in closer systems? What is the
distribution of galaxy asymmetries for the local universe? Are there
isolated galaxies with high asymmetry, and are these merger remnants?

\cite{patton05} carried out a similar study by comparing the
asymmetries of isolated galaxies and those in close pairs, taken from
the CNOC2 redshift survey, using Hubble Space Telescope snapshots of
close pair candidates and a small sample of isolated CNOC2 galaxies
falling within the HST field of view. For these galaxies, they measure
asymmetries using the $R_T+R_A$ index of \cite{schade95}. Galaxies in
CNOC2 pairs are more asymmetric than isolated ones, but over only a 
limited range of pair separations (up to $21\ h^{-1}$~kpc 
in projection) set by the original pair selection by \cite{patton02}. 
\cite{hernandez05} have compared asymmetries in spiral-spiral pairs, 
ultraluminous infrared galaxies (commonly regarded as interacting and/or 
merging systems; \citealt{lonsdale06}) and a local sample of objects from 
\cite{frei96}, showing that asymmetry increases with decreasing pair separation, 
although the samples were somewhat heterogeneous.

To measure the merger rate from asymmetry {\em and} dynamically close
pairs requires a dataset with both high-quality imaging {\em and} very
complete redshift information. The Millennium Galaxy Catalogue (MGC;
\citealt{liske03}) is particularly well-suited to this task. The MGC
consists of deep CCD imaging of a long, thin ($36' \times 72^{\circ}$)
equatorial strip, reaching to $\bmgc=24$~mag and $\mu_B=26$~mag
arcsec$^{-2}$, and coinciding with both the 2dF Galaxy Redshift Survey
northern strip (2dFGRS; \citealt{colless01}) and with the Sloan
Digital Sky Survey data release 1 region (SDSS; \citealt
{abazajian03}). The MGC also includes a parallel redshift survey,
reaching $99.8$\% completeness at $\bmgc=19.2$~mag and $96.0$\% at
$\bmgc=20$~mag \citep{driver05}.

Because of its homogeneous high-quality imaging, its large, contiguous
field of view, and high redshift completeness, the MGC enables one to
compute the merger rate from the statistics of dynamically close pairs
\citep{depropris05} {\it and} galaxy asymmetries. The main purpose of
this paper is to carry out a comparison of these two approaches using
the same, homogeneous, sample of galaxies. As in previous MGC papers,
we adopt a cosmology with ${\Omega_M} = 0.3$, $\Omega_{\Lambda} = 0.7$
and calculate all distances with reference to $H_0 = 100\
h$~km~s$^{-1}$~Mpc$^{-1}$.

\section{Methodology}

\subsection{Sample}

\begin{figure}
\plotone{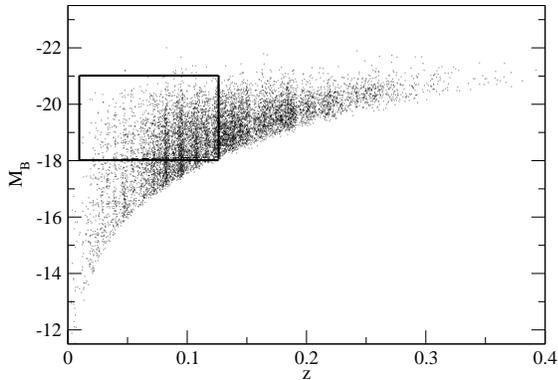}
\caption{The volume-limited box used to select galaxies for analysis.
The points are all MGC galaxies with $\bmgc < 20$~mag.}
\label{selection}
\end{figure}

In order to compare asymmetries and close pairs as indicators of the
merger rate, we first select a volume-limited sample of galaxies from
the MGC. This sample includes $3237$ galaxies with $\bmgc < 20$~mag,
$-21 < M_B - 5 \log h < -18$~mag and $0.01 < z < 0.123$. These limits
are shown in Figure~\ref{selection}. From this sample we extract close
pairs with projected separations of up to $100\ h^{-1}$ kpc and isolated
galaxies (i.e., with no companion within the specified magnitude and redshift
limits and projected separation criteria). We also compute asymmetries for
all galaxies in the volume-limited sample. Because of the absolute magnitude
range considered, we can identify mergers with a luminosity ratio of up to
$1:16$ for the brighter galaxies ($M_B=-21$ mag.), with the sample becoming more
incomplete for fainter galaxies. The range in luminosity ratios that can be
explored biases the sample towards pairs with nearly equal luminosities.

\subsection{Galaxy Pairs}

We identify dynamically close pairs using the above sample in the same way 
as \cite{patton00,patton02} and \cite{patton05}. However, we consider a 
wider range of projected separations than \cite{patton05} because of our 
wider angular coverage and more complete spectroscopy. Two galaxies in our 
sample form a dynamically close pair if their projected separation is less 
than $100\ h^{-1}$~kpc, and their velocity difference is less than $500$~km
~s$^{-1}$. We consider only the {\it closest} companion (multiplets are ignored) 
and galaxies with no such companion are regarded as isolated for our purposes.

The above criteria for position and velocity separation were chosen as
follows. The pairwise velocity dispersion of 2dF galaxies is $\sim
500$~km~s$^{-1}$ \citep{hawkins03}; therefore objects with larger
velocity separations are unlikely to be bound (see also Figure~4 in
\citealt{patton00}). In addition, simulations show that pairs with
larger velocity separations are unlikely to merge (Carlberg, Pritchet
\& Infante 1994), while a velocity separation of $500$~km~s$^{-1}$ is
generally larger than the internal velocity dispersion of the galaxies
under study, which decreases the likelihood of a significant
interaction \citep{makino97}.

Simulations also show that galaxies with similar masses, and with a
projected separation of less than $20\ h^{-1}$~kpc are likely to merge
within $1$~Gyr \citep{barnes88,carlberg94, conselice06}. The upper
limit on the projected separation we adopt in this paper ($100\
h^{-1}$~kpc) is arbitrary, but it should isolate all objects likely to
merge in the near future, as well as those where the interaction has
induced significant asymmetry (cf.\ \citealt{hernandez05}).

It is of course possible that we miss some companions due to our
incomplete redshift coverage. There are $399$ $\bmgc < 20$~mag
galaxies (4\% of all $\bmgc < 20$ systems) in the MGC for which we have 
not obtained a redshift. We use the projected separation criterion to 
calculate whether any of these galaxies could be the closest companion to a galaxy in the
volume-limited sample. If so, we assume that the galaxy without
redshift has the same redshift and $k$-correction as its putative
companion, and require that it obeys the absolute magnitude cuts to be
part of the volume-limited sample. Objects in the sample for which the
closest companion may be one of the $399$ galaxies without redshifts
are however not included in our analysis. This decreases our sample to $3184$
galaxies from the original $3237$. This contains $112$ galaxies in
pairs with projected separation of less than $20\ h^{-1}$~kpc, and
$2561$ isolated galaxies. While all pairs within the specified 
luminosity and redshift limits are included, the sample is incomplete
towards pairs with large luminosity differences and tends to be 
more complete for pairs of nearly equal luminosities. The median
magnitude difference for the above 112 pairs is $0.6$ mag., equivalent
to a typical 1:2 luminosity ratio (i.e., we primarily measure
major mergers).

\subsection{Galaxy Asymmetries}
\label{asym}

Asymmetries for galaxies within the volume-limited sample are 
calculated using both the $R_T+R_A$
index of \cite{schade95} and the more recent CAS indices of
\cite{conselice00a}, which have been calibrated on local galaxy samples
to identify merger remnants. The details on how to calculate the
asymmetries used in this paper are given in \cite{simard02} and
\cite{allen06}, for the $R_T+R_A$ method, and \cite{conselice00a,
bershady00,conselice03,conselice04} for the revised CAS indices. 
The CAS (concentration, asymmetry, clumpiness) parameters are a 
non-parametric method for measuring the structures of galaxies resolved 
on CCD images. One particular feature of this system is that galaxies 
with extremely high asymmetries, usually $A > 0.35$, are likely to be 
in a major (post-)merger phase. This has been calibrated with asymmetry
measurements of nearby and distant normal and merging galaxies
\citep{conselice00a,conselice03,conselice05,bridge07}, and of `galaxies' 
seen in N-body merger simulations \citep{conselice06}. Furthermore, 
Conselice, Bershady \& Gallagher (2000b) have shown that galaxy mergers 
have both high asymmetries and high HI line width asymmetries.

Our method for computing the CAS parameters is slightly different from
that in previous works. Because we are examining nearby bright galaxies in
a general field survey, our sample was not
specifically selected to be clean of foreground and background
contamination.  Our sample therefore represents a more generalised
case for measuring asymmetries in field surveys of nearby galaxies.  For
each galaxy we use a SExtractor \citep{bertin96} segmentation map to
replace all nearby objects with the sky, using the same noise
properties as measured directly from the background. In this way we
eliminate the possible contamination of the asymmetry measurement by
these objects. This method is however not always perfect when there
are very nearby bright stars or large galaxies. We must therefore keep
in mind that any sample of highly asymmetric galaxies may be
contaminated by galaxies that have an intrinsically low asymmetry, but whose
asymmetry measurement was corrupted by a nearby object.

\section{Results}

\subsection{The Asymmetry Distribution}

\begin{figure}
\plotone{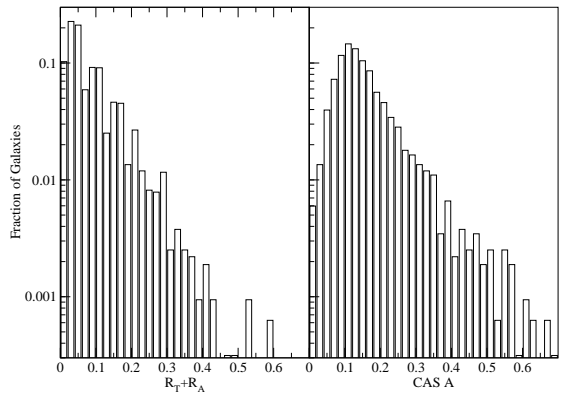}
\caption{The distribution of asymmetry values for all galaxies in the 
volume-limited sample. Note that the ordinate is plotted logarithmically
to better show the small number of asymmetric galaxies.}
\label{asym_distr}
\end{figure}

Figure~\ref{asym_distr} shows the distribution of asymmetries for both
the \cite{schade95} and CAS methods: we plot the fraction on the
ordinate logarithmically, in order to emphasise the small number of
more asymmetric galaxies. 

For the $R_T+R_A$ index we find that the asymmetry distribution peaks
at $\sim 0.05$, with a long tail of more asymmetric objects. Following
the definitions of \cite{patton05}, ($18.8 \pm 0.8$)\% of galaxies are
`asymmetric' ($R_T+R_A \geq 0.13$) and ($12.4 \pm 0.7$)\% are `highly
asymmetric' ($R_T+R_A \geq 0.16$). However, unlike the CAS indices, 
$R_T+R_A$ has not been calibrated to specifically identify merging
systems or merger remnants. Visual examination of the $2$\% most 
asymmetric objects (by the $R_T+R_A$ criterion) shows that about
1/3 are in close pairs, 1/3 seem to be merger remnants and the rest
appear to be edge-on galaxies or classical irregulars and late-type
spirals (see also section below). 

Among the close pairs with projected separation of less than $20\
h^{-1}$~kpc we find that ($25.5 \pm 5.5$)\% are `asymmetric' and
($16.9 \pm 4.3$)\% are `highly asymmetric'. The pair fractions and
merging pair fractions (assuming that 50\% of the pairs actually
merge) are in good agreement with the values derived by \cite{patton05},
but we find a much larger fraction of asymmetric or highly asymmetric
isolated galaxies. This is likely due to the relatively small sample
of isolated galaxies in \cite{patton05}, the lower physical resolution
of the comparatively shallow HST imaging used (leading to a smoothing
of the low surface brightness brightness asymmetric features), different
isolation criteria used and treatment of segmentation maps.

The CAS method defines galaxies with $A > 0.35$ as unusually
asymmetric: \cite{conselice00a} suggest that such objects may be
interpreted as merger remnants. The distribution of CAS $A$ values in
Figure~\ref{asym_distr} peaks at $\sim 0.11$, suggesting that $A$ is
sensitive to relatively minor deviations from symmetry. We find that
($4.1 \pm 0.4$)\% of our sample have $A > 0.35$ with this 
blind selection. At face value, and assuming that $50$\% of the close 
pairs (with separation less than $20\ h^{-1}$~kpc) merge, this is about 
a factor of two larger than the `merging pair' fraction. However, before 
we draw any conclusions regarding possible discrepancies between the two 
methods we will need to evaluate more carefully what types of objects 
have been selected by the $A > 0.35$ cut.

\begin{figure}
\plotone{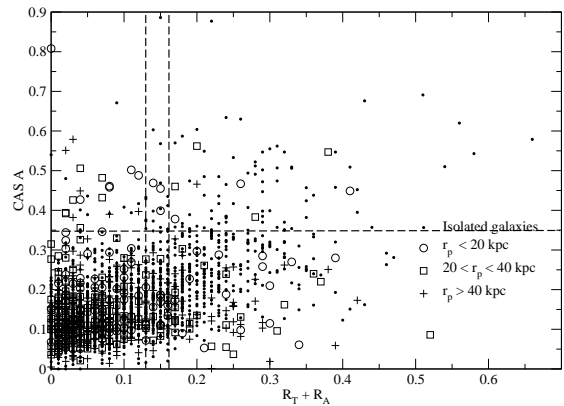}
\caption{Comparison of CAS $A$ values and $R_T+R_A$ for galaxies
in different separation classes. The horizontal thick dashed line
indicates the $A=0.35$ asymmetry limit in the CAS system. The two
vertical thick dashed lines show the $R_T+R_A=0.13$ and $0.16$
limits.}
\label{comp}
\end{figure}

Figure~\ref{comp} compares asymmetry measurements for galaxies in various
isolation classes (from close pairs to galaxies we regard as being isolated)
for both CAS $A$ and $R_T+R_A$ and shows the limits used to distinguish asymmetric
or highly asymmetric galaxies for both methods. Not surprisingly, $R_T+R_A$
thresholds appears to be considerably less stringent than CAS $A$, although there is a 
broad relation between the two indices. One can see that there are several
galaxies in close pairs with separations less than $40\ h^{-1}$ kpc above
the $A=0.35$ line, as well as numerous isolated galaxies, while only a few
pairs with larger separation have high CAS asymmetry. There are numerous
objects deemed as highly asymmetric by the $R_T+R_A$ index, but with $A 
 < 0.35$. In most cases, these are edge-on or late-type spirals which have
not been decontaminated from the sample as we do for the high $A$ systems.

\subsection{The Nature of Asymmetric Galaxies}
\label{natasym}

A blind study such as this requires that we examine by eye those systems which 
are asymmetric.  There are two reasons why one should not simply assume that 
{\em all} MGC galaxies with $A > 0.35$ are imminent mergers or merger remnants.
First of all, as pointed out in Section \ref{asym}, the sample may contain objects 
whose asymmetry measurements have been corrupted by nearby bright/large objects. 
Secondly, the $A$ value of $0.35$ \citep{conselice03} was established using the 
sample of normal nearby galaxies of \citet{frei96} and a collection of starburst 
galaxies, which were selected to span a large range of morphological types. In 
contrast, here we are using a volume-limited sample. Also, although the CAS $A$ 
is reasonably robust, it does depend on the physical resolution of the data
\citep{conselice00a,conselice03}. Therefore we have visually examined all $129$
galaxies with $A > 0.35$ in our sample and classified them into the following 
categories:

\begin{itemize}

\item{\sc Likely merger remnants}: although this is a somewhat
subjective classification, these objects appear to be relatively
obvious `trainwrecks' or very disturbed disks with multiple, similarly
bright light components. We find $23$ objects in this category,
and we show MGC postage stamp images of these galaxies in
Figure~\ref{mmr}.

\begin{figure*}
\plotone{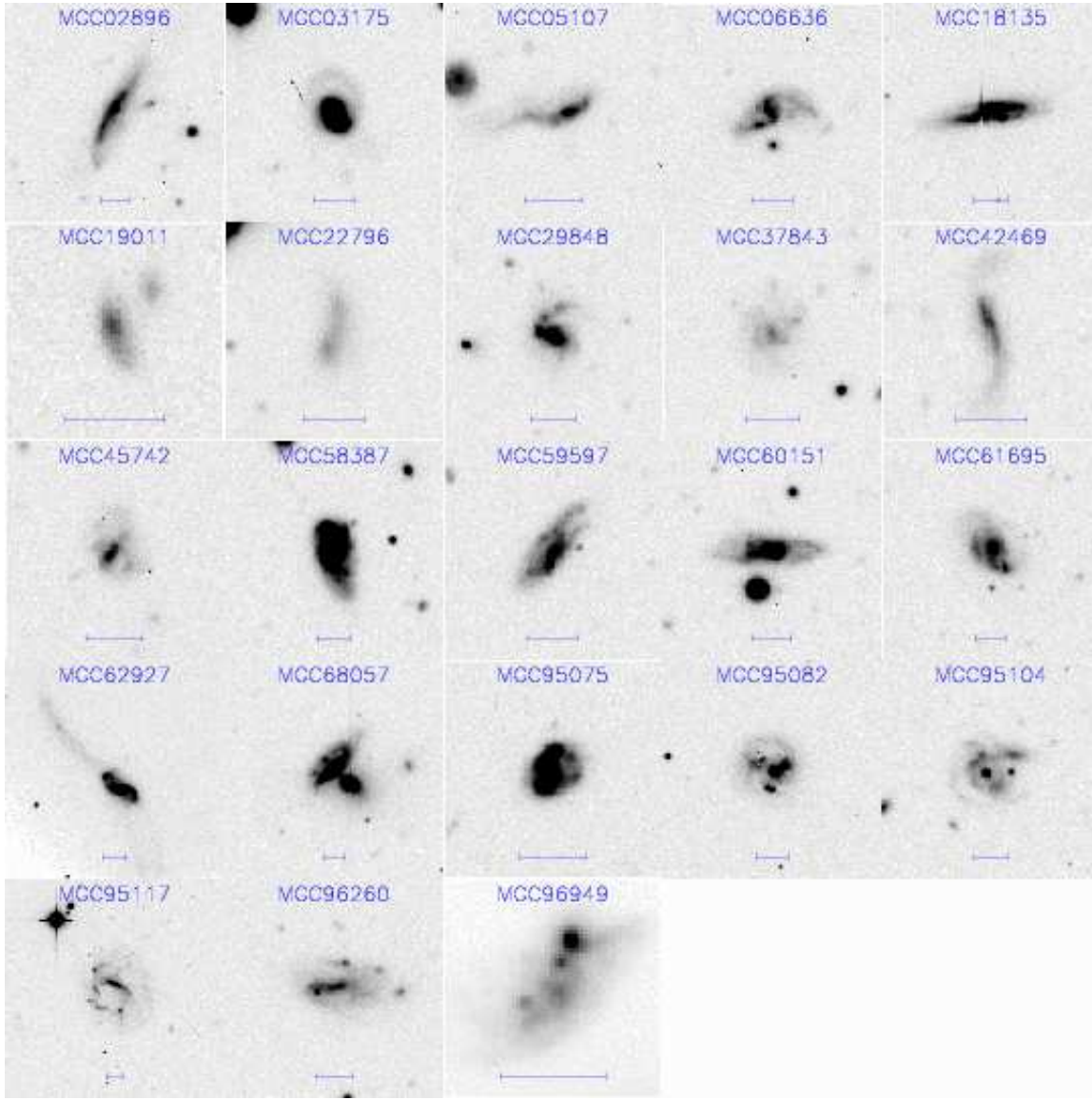}
\caption{Postage stamp MGC images of galaxies with $A > 0.35$ which
are regarded as merger remnants. The scale of each image is indicated
by the horizontal bar which is $10$~arcsec.}
\label{mmr}
\end{figure*}

\item{\sc Likely imminent mergers}: these are disturbed galaxies
similar in appearance to the above, but satisfying two additional
requirements: (i) they are a member of a close pair as identified
above, i.e.\ have a `major' companion; (ii) they show evidence of
interaction with their companion. We find $13$ such objects 
and show their images in Figure~\ref{imm}. The projected
separation to the companion, $d$, in this sample ranges from $2.5$ to
$34\ h^{-1}$~kpc. 

\begin{figure*}
\plotone{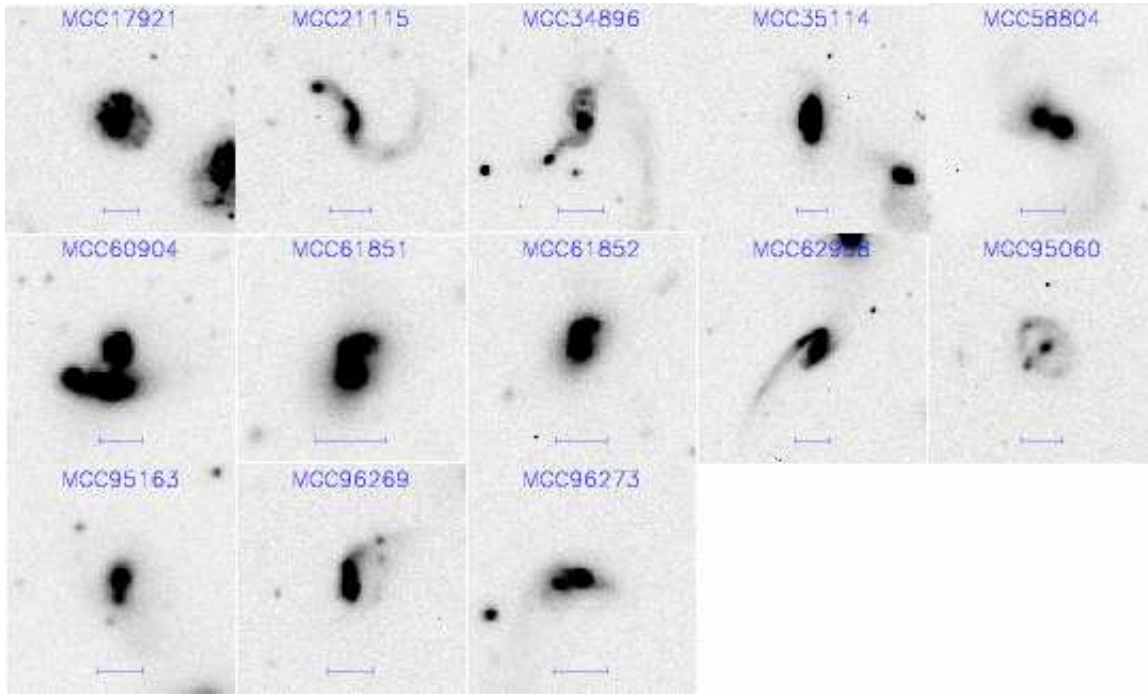}
\caption{As Figure~\ref{mmr} but for galaxies with $A > 0.35$ regarded
as imminent mergers.}
\label{imm}
\end{figure*}

\item{\sc Other}: there are 8 galaxies that are not imminent mergers and 
whose asymmetry measurements are not contaminated (see below). However, 4 
of these galaxies are faint and/or low surface brightness objects which we 
cannot reliably classify, and all of these look unusual and could be merger
remnants. In contrast, the other 4 galaxies in this category are most
likely not recent merger remnants: one is a face-on disk, one is a
spheroidal galaxy (with some low-level debris nearby), and the other two
are {\em minor} mergers (as determined from the full MGC catalogue).

\begin{figure*}
\plotone{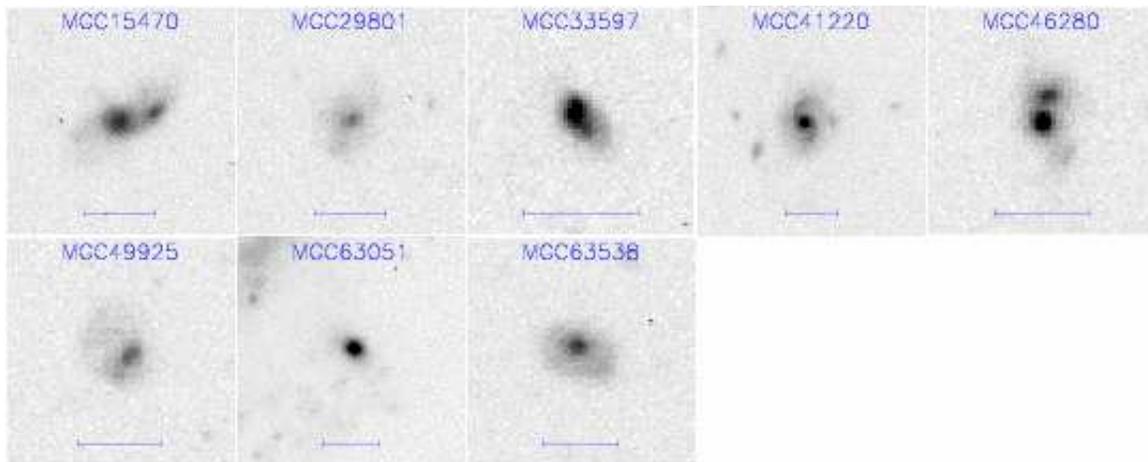}
\caption{As Figure~\ref{mmr} but for galaxies with $A > 0.35$ which cannot
be reliably classified.}
\label{oth}
\end{figure*}

\item {\sc Contaminated:} galaxies where the asymmetry appears to be
due to the presence of a bright star in the vicinity. $28$ galaxies
are thus affected. Furthermore, for 57 galaxies SExtractctor did not 
construct correct segmentation maps, which led to erroneously high 
CAS $A$ values. If these objects are (wrongly) included in the sample 
of highly asymmetric galaxies they account for 50\% of the sample. 
Hence, without paying particular attention to the construction of the 
segmentation maps, and/or visual vetting of the sample, a large fraction 
of $A > 0.35$ galaxies turn out to be false positive mergers. 

\end{itemize}

If we exclude objects with contaminated measurements and with bad segmentation 
maps, we find that $\sim 80$\% of the asymmetric galaxies in our sample are obvious 
imminent or recent mergers.  The status of the remainder is more uncertain. However
of the remaining 8 systems, four are possible merger remnants.

We note that over half of all systems initially found with $A > 0.35$ 
were contaminated by stars, or were over-segmented. This demonstrates 
that vetting of asymmetry selected merger samples is advisable (see also 
\citealt{conselice03}). Although contamination by stars is less of a problem 
at high redshifts, where, presumably, the asymmetry index will be applied 
in future surveys, contamination by neighbouring unrelated galaxies may be 
an issue \citep[e.g.][]{kampczyk07}. 

As noted above, we found from our visual inspections that $57$ of our
initial sample of $A > 0.35$ galaxies were asymmetric due to Sextractor
having produced bad segmentation maps for these objects. Most of these
were late-type, face-on or edge-on spirals with prominent star-forming
regions and/or spiral arms, and SExtractor falsely separated these into
multiple sources. As noted in Section 2.3, the CAS code attempts to avoid
the corruption of a galaxy's CAS parameters through nearby objects by
replacing these with blank sky. In the case of the shredded galaxies this
procedure created `holes' in their light distribution and hence
artificially high asymmetries. To deal with this we re-ran the CAS code on
all galaxies without using any segmentation maps at all. Each galaxy was
then assigned the smaller $A$ value from the two runs, and $57$ galaxies 
went from $A > 0.35$ to $A < 0.35$.

Hence, in order to avoid significant contamination of high-asymmetry 
samples with late-type disks in future large surveys (where visual 
inspection will be unfeasible), care must be taken when constructing the 
segmentation maps and/or when masking neighbouring objects.

\subsection{Asymmetry and Pair Separation}
\label{asymps}

\begin{figure}
\plotone{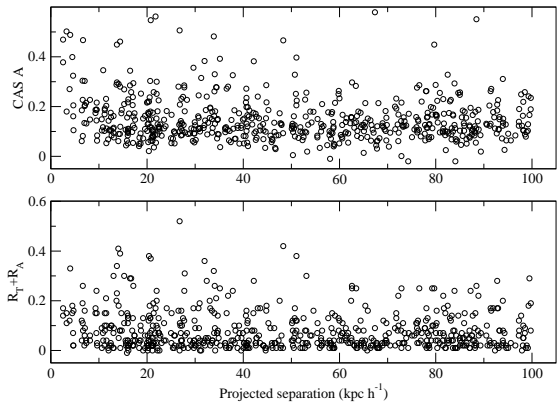}
\caption{Galaxy asymmetry versus pair separation using both the
$R_T+R_A$ and the CAS $A$ indices.}
\vspace{1cm}
\label{asym_pair}
\end{figure}

Figure~\ref{asym_pair} shows the values of $R_T+R_A$ and CAS $A$ for
galaxies in pairs, plotted against the projected separation of the
pair members. Galaxies in pairs with $d < 50\ h^{-1}$~kpc clearly show
an excess asymmetry over galaxies in pairs with wider separations.  This
appears to confirm the findings of \cite{hernandez05} and
\cite{patton05} that being in a close pair induces asymmetries (via
tidal stresses and star formation). Hence there is no doubt that at
least a fraction of the close pairs consist of physically bound
objects undergoing an interaction.

\begin{figure}
\plotone{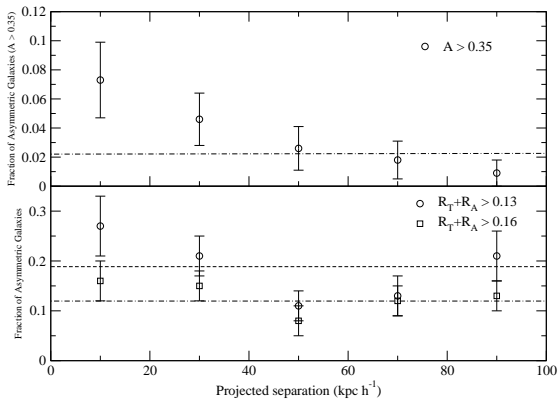}
\caption{Fraction of asymmetric galaxies in five projected separation
bins. The horizontal lines show the fraction of isolated galaxies with
$R_T+R_A > 0.13$ (dashed line) or $>0.16$ (dot-dashed line) or $A >
0.35$ (dashed line in upper panel).}
\label{asym_frac}
\end{figure}

In order to show this more clearly we plot the fraction of `asymmetric'
galaxies vs.\ pair separation for five $20\ h^{-1}$~kpc wide bins of
projected separation in Figure~\ref{asym_frac}. We also indicate the
fraction of asymmetric galaxies (according to both methods) among the
isolated galaxies. The fraction of asymmetric galaxies increases with
decreasing pair separation, consistent with the interpretation that
closer pairs result in a progressively stronger interaction.

\begin{figure*}
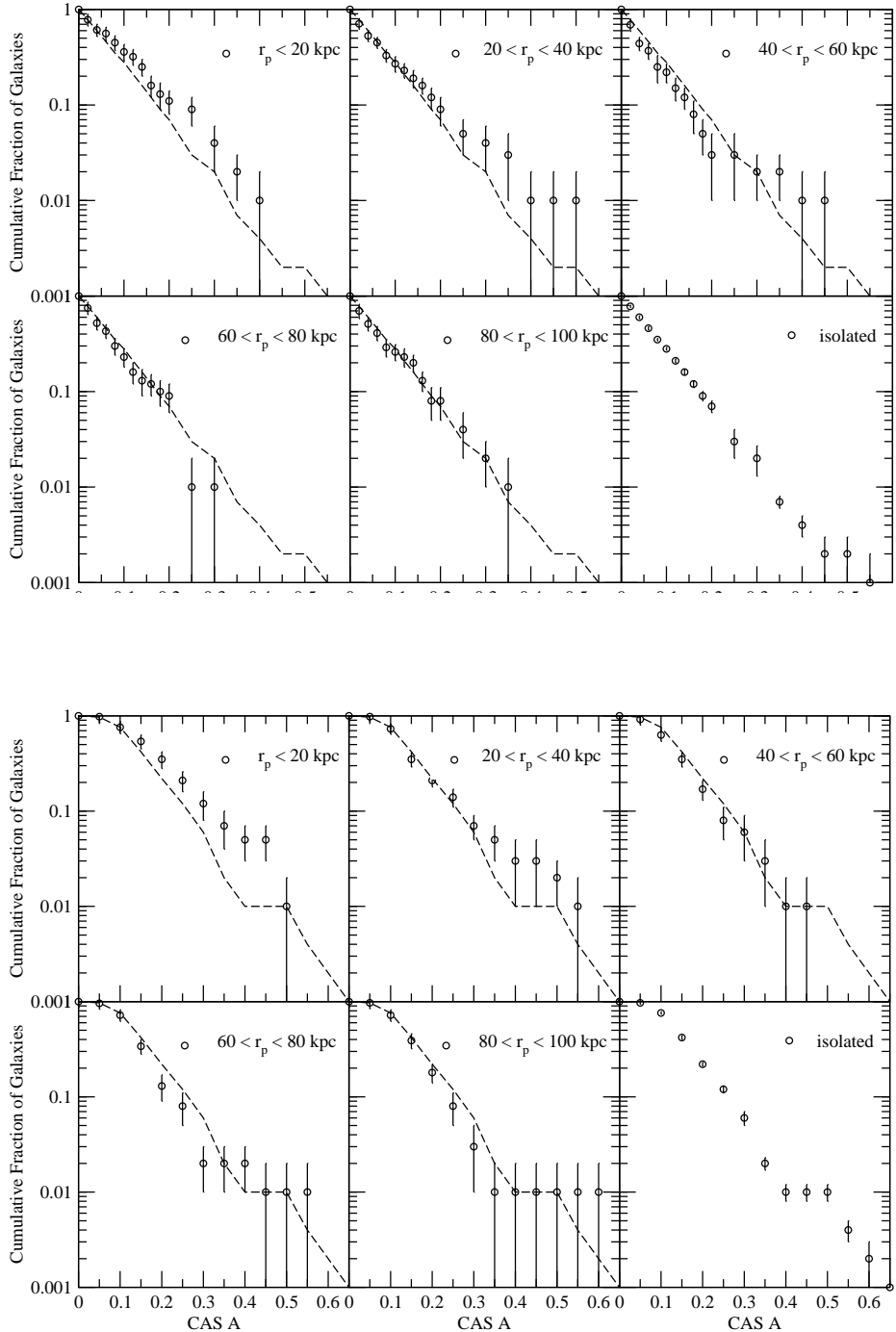

\epsscale{.80}
\plotone{f8a.eps}\\
\vspace{1cm}
\plotone{f8b.eps}
\caption{Cumulative fraction of galaxies with asymmetry greater than
shown on the horizontal axis. The dashed line in each panel shows the
observed asymmetry distribution for isolated galaxies (shown in the
bottom right-hand panel).}
\label{asym_distr_d}
\end{figure*}

We also compare the cumulative distribution of asymmetries for the
five projected separation bins used above, and for isolated galaxies
in Figure~\ref{asym_distr_d}. As implied by Figures~\ref{asym_pair}
and \ref{asym_frac}, we see an excess of asymmetric galaxies over the
distribution for isolated galaxies only in the two lowest separation
bins. However, in the smallest separation bin, there is an excess
fraction of asymmetric galaxies even at the highest asymmetries.
This implies that, at least in the closer pairs, we are witnessing a
significant degree of morphological disturbance.

While these figures show that $R_T+R_A$ is also capable of identifying
mergers and merger remnants, this index has not been calibrated with the
same thoroughness as the CAS index, where extensive work has been carried 
out by Conselice and collaborators. For this reason, from now on, we use the CAS
index exclusively to derive the merger rate and compare to the pair fractions.

By comparing the differential CAS $A$ distribution of isolated galaxies
with that of galaxies in close pairs with $d < 20\ h^{-1}$~kpc we can
derive a limit on the fraction, $f$, of unphysical pairs in this sample.
For our purposes, we assume (or define) galaxies in `unphysical' pairs 
to be those that have the same $A$ distribution as isolated galaxies.
This calculation will give us therefore a lower limit on the fraction of 
paired galaxies that are in interacting systems. We denote the normalised 
differential $A$ distribution of isolated galaxies by $i(A)$, and the unknown 
distribution of galaxies in true, physical pairs as $t(A)$. Hence the 
normalised $A$ distribution of {\em all} apparent galaxy pairs 
with $d < 20\ h^{-1}$~kpc can be written as: $p(A) = f \; i(A) + (1-f) \; t(A)$. 
Since $f$ must lie in the range $0 \le f \le 1$ and since $t(A) \ge 0$ for 
all $A$, we find that $p(A) \ge f \; i(A)$ must hold for all $A$. By comparing 
the observed distributions $p(A)$ and $i(A)$ we find that the maximum value of $f$
for which this condition still holds is $0.65$, where this constraint
is driven by the low-$A$ end of the distributions ($A < 0.15$) as
expected. In other words, of all galaxies in dynamically close pairs
with $d < 20\ h^{-1}$~kpc at least $35$\% lie in physical pairs. This
compares well with the fraction of $50$\% estimated by
\citet{patton00}, and we will use this number in the following
calculations.

\section{Pair and Asymmetry Merger Time-scales: Implications for the
Merger Fraction and Galaxy Evolution}

We are now in a position to calculate the merger time-scale and
fraction for close pairs and asymmetry-selected galaxies, and compare
the lifetime of the merger remnant to the merger time-scale for the
progenitor pair.

In Section \ref{asymps} we found that only pairs with small separation
($d < 50\ h^{-1}$~kpc) are likely to merge within a short time interval, 
and that among these the
majority of likely mergers are found among pairs with $d < 20\
h^{-1}$~kpc (based on the presence of asymmetries as an indicator of
an on-going interaction). To a first approximation the close pair
fraction for these latter objects can give a measure of the merger
rate. We find that the pair fraction is ($3.5 \pm 0.4$)\%. In order to
correct for pairs which are missed because of redshift incompleteness
we adopt the $18.5\%$ incompleteness correction calculated by
\cite{depropris05} for MGC close pairs with $d < 20\ h^{-1}$~kpc,
yielding a close pair fraction of ($4.1 \pm 0.4$)\%. Assuming that
$50$\% of these close pairs are likely to be physical pairs and hence
future mergers, we derive a merger fraction of ($2.1 \pm 0.2$)\% for
our volume-limited sample.

In Section \ref{natasym} we found that high asymmetry identifies both
merger remnants and to a lesser extent future mergers. To derive a
merger fraction that is directly comparable to the one derived from
the close pairs above (which is essentially a `progenitor galaxy'
fraction), we must take into account that each merger remnant was
produced by (at least) two progenitor galaxies and hence count it
twice. In addition we apply the above incompleteness correction to the
$8$ objects in pairs with $d < 20\ h^{-1}$~kpc. This yields a galaxy
merger fraction of ($1.9 \pm 0.2$)\%.

The similarity between the merger fractions derived by these two
methods suggests that the total time-scale for asymmetry (i.e.\
including the pre-merger phase) is similar to the merger
time-scale of a $d < 20\ h^{-1}$~kpc pair. We can determine the 
pre- and post-merger fraction of the total asymmetry time by comparing
the number of
imminent mergers (corrected for redshift incompleteness) with the
number of merger remnants ($\times 2$) identified by asymmetry. We
deduce that the pre-merger phase takes up $24$\% of the total
asymmetry time, consistent with results derived from N-body
simulations (Conselice 2006).
We notice, however, that there are several galaxies in pairs with
$A > 0.35$ that have large separations (some as large as 34 $h^{-1}$
kpc). It is likely that these objects are on parabolic or hyperbolic
orbits or may not merge until after several more passages.

The general picture that emerges from these results is the following:
close pair selection (with $d < 20\ h^{-1}$~kpc) is sensitive to most
stages of the `pre-merger'. From dynamical friction considerations
(see equation 7 in \citealt{conselice06}) we find that the merger
time-scale for a `typical' close pair in our sample, with average
separation of $12\ h^{-1}$~kpc and average velocity separation of
$150$~km~s$^{-1}$ is about $0.3$~Gyr. Being in a close (physical) pair
enhances asymmetry and during the last $\sim 22$\% (or $65$~Myr) of
this pre-merger phase the asymmetry has become so significant that the
imminent merger is selected by the $A > 0.35$ cut. The total asymmetry
time is much longer than this because the remnant remains asymmetric
for $0.21$~Gyr after the two systems merge, and the new system
dynamically relaxes \citep{conselice06}. Hence, as implied by N-body
simulations, asymmetry identifies the immediate pre-merger stage as
well as an extended post-merger phase. While the absolute time-scales
above are quite uncertain their ratios are not, but are fixed by our
results, i.e., if the merger time-scale is revised to $1$~Gyr then the
asymmetry time-scales must also change accordingly.

A comparison with theoretical predictions is somewhat difficult, as
there may not be a one-to-one correspondence between galaxies and
their host dark haloes (see, e.g., \citealt{berrier06}). With this
caveat, \cite{maller05} predict a merger rate of $0.054$ mergers per
Gyr for massive galaxies with mass ratio 1:1 to 1:3. From our data we
find that $66$\% of the MGC pairs consist of galaxies within this mass
ratio, yielding a pair fraction of $\sim 0.014$. The merger rate we
derive is marginally consistent with the above prediction for a merger
time-scale of $0.3$~Gyr or too low by a factor of $3.9$ for a
time-scale of $1$~Gyr, similar to the comparison shown by Bell et
al. (2006 -- their figure 2) from theoretical models. 

The merger rate we derive, following equation 3 of \cite{lin04}, is $(5.2 \pm 1.0) 
\times 10^{-4}$ $h^3$ Mpc$^{-3}$ Gyr$^{-1}$. Strictly speaking this is a lower
limit, as some pairs are certainly missed, but it should be a fair estimate of
the major merger rate for nearly equal luminosity pairs, which are less incomplete.
For comparison, \cite{lin04} derive an average merger rate of $4 \times 10^{-4}\ h^3\ 
{\rm Mpc}^{-3}\ {\rm Gyr}^{-1}$ at $0.5 < z < 1.2$ for galaxies with $-21 < M_B < -19$, 
which is about one order of magnitude lower than that derived by \cite{conselice03},
but in good agreement with our $z \sim 0.1$ value. It is much larger, by almost two
order of magnitude, than the merger rate between luminous red galaxies at $z < 0.36$
measured by \cite{masjedi06}, suggesting that the majority of mergers between galaxies
in the $z < 0.5$ universe are not `dry' \citep[c.f.][]{blanton06}. Our results are 
consistent with a flat evolution of the merger rate, i.e., as $(1+z)^{\sim 1}$ \citep{lin04}
rather than the $(1+z)^3$ dependence expected from CDM models (\citealt{gottlober01}, 
but see discussion by \citealt{berrier06}). If the merger rate evolution is flat out to 
$z \sim 1$ only a small fraction of massive galaxies can have formed via major mergers 
in the last 1/2 of the Hubble time.

\section{Summary}

In this paper we have compared the close pair fraction and the
fraction of asymmetric galaxies in a volume-limited sample of galaxies
drawn from the MGC. Our sample consists of $3184$ galaxies between
$-21 < M_B - \log h < -18$~mag and $0.01 < z < 0.123$. The main
conclusions of this work are:

\begin{itemize}

\item The CAS selection for mergers, $A > 0.35$, after correction for 
contamination,
is highly successful at locating galaxies in major mergers with
a corrected fraction of $\sim 80$\%.   The use
of a blind selection however can result in many false positives.
To create a clean merger sample selected with the
asymmetry index it is necessary to either visually examine all
high-$A$ galaxies, or use other automated methods, such as a high
clumpiness \citep{conselice03}, and axis ratios to remove
contamination.

\item We find a connection between pairs and asymmetry. Galaxies in
pairs tend to be more asymmetric and the asymmetry tends to increase
with decreasing pair separation, implying that these galaxies are
actually interacting. We estimate that at the very least $35$\% of galaxies in
close pairs with projected separation of less than $20\ h^{-1}$~kpc
are actually physically bound based on their excess distortion
compared to isolated galaxies.

\item The overlap between our close pair and asymmetry-selected merger
samples is small. Of all galaxies in pairs with projected separation
less than $20\ h^{-1}$~kpc only $7$\% were identified as likely merger
events by the asymmetry method. Hence it is likely that the pair and
asymmetry methods are tracing somewhat different phases and perhaps
mass ratios of merging galaxies, while some objects in pairs are not
asymmetric enough to be tagged by the CAS indices.

\item We find that after correcting the close pair sample for chance
superpositions and the asymmetry sample for contamination (by stars,
and most importantly bad segmentation maps) that 
the merger fraction and rate for the two methods are
very similar. This suggest that both methods are tracing the same
underlying merger process, although at somewhat different phases of
the merger process.

\item Once we have accounted for contamination and incompleteness in
both methods we derive a consistent merger fraction of $1.9\%$ per $0.3$~Gyr
and a merger rate of $(5.2 \pm 1.0) \times 10^{-4}$ $h^3$ Mpc$^{-3}$
Gyr$^{-1}$. 

\end{itemize}

\section*{Acknowledgements}

The Millennium Galaxy Catalogue consists of imaging data from the
Isaac Newton Telescope obtained through the ING Wide Field Camera
Survey programme. The INT is operated on the island of La Palma by the
Isaac Newton Group in the Spanish Observatorio del Roque de los
Muchachos of the Instituto de Astrof{\'i}sica de Canarias.
Spectroscopic data come from the Anglo-Australian Telescope, the ANU
2.3m, the ESO New Technology Telescope, the Telescopio Nazionale
Galileo and the Gemini North Telescope. The survey has been supported
through grants from Particle Physics and Astronomy Research Council
(UK) and the Australian Research Council. RDeP and CJC also
acknowledge grants from PPARC at the University of Bristol and at the
University of Nottingham. DRP gratefully acknowledges support from the
Natural Sciences and Engineering Research Council of Canada. The data
and data products are available from
http://www.eso.org/$\sim$jliske/mgc or on request from JL or SPD.

\setlength{\bibhang}{2.0em}

\end{document}